\documentclass[a4paper,11pt]{article}
\usepackage{caption}
\usepackage{algorithmic}
\usepackage{marvosym}
\usepackage{tipa}
\usepackage{algorithm}
\usepackage{mathtools}
\usepackage{eurosym}
\usepackage{amsfonts}
\usepackage[margin=2cm]{geometry}
\usepackage{graphicx}
\usepackage{tikz}
\usepackage{hyperref}
\usepackage{amsmath}
\usepackage{pgfplots}
\title{\bf {\sc Merchant Sharing}\\$\color{red}{\mathcal{T}_i^\nearrow{}=i\mathcal{P}_i^\searrow{}}$\\{\small Towards a Zero Marginal Cost Economy}}
\author{Laurent {\sc Fournier} -- {\small\{\url{laurent.fournier@cupfoundation.net}\}}}

\begin{document}

\usetikzlibrary{shapes,fit,arrows,shadows,backgrounds,svg.path,mindmap,trees}
\maketitle
\begin{abstract}
This paper is the first attempt to formalize a new field of {\em economics}; studding the {\em Intangibles Goods} available on the {\em Internet}.
We are taking advantage of the {\em digital world}'s specific rules, in particular the {\bf zero marginal cost}, to propose a theory of {\sc trading} \& {\sc sharing} unified. A function based money is created as a world-wide currency; $\sqcup$ (pronounced \textipa{/k2p/}). We argue that our system discourage speculation activities while it makes easy captured taxes for governments. The implementation removes the today's {\em paywall} on the {\em Internet} and provides a simple-to-use, open-source, free-of-charge, highly-secure, person-to-person, privacy-respectful, digital payment tool for citizens, using standard smart-phones with a strong authentication. Next step will be the propagation of the network application and we expect many shared benefits for the whole economics development. 
\vspace{.5em} \\
\noindent {\bf keywords:} {\footnotesize Economics, Internet, Intangible Good, Sharing, Trading, Money, Digital Signature, Payment, Currency, Exchange Rate, Cultural Piracy, Copyright, Paywall, Peer-to-peer, Zero Marginal Cost, Collaborative Commons, Speculation, Open-source.}
\end{abstract}
\section{Introduction}
{\sc Sharing} and {\sc Trading} seem first conflicting in our everyday physical experience of {\em Tangibles Goods} ({\sc tg}). When we share something, it's usually free of charge and there is no declared ownership. People are involved in a strong relationship, as friends. On the contrary, buying an object defines explicitly the owner and exclude other people from having or using the good. Trading means that the seller is dispossessed of the good against a financial reward, after a {\sc one-to-one} instantaneous relationship. Furthermore, the buyer and the seller may be completely anonymous from each other.\\
 Is a {\em Merchant Sharing Theory} likely to be impossible?\\
Well, the {\em digital} world carried by the {\em Internet} is following different rules than the physical world. Our purpose here is to show that over the {\em Internet}, {\sc sharing} and {\sc trading} are not only compatibles, but have tremendous advantages to be associated.
The following theory introduces a breakthrough in the {\bf economics} domain, making new potentially growing markets and new business opportunities. Since {\em Internet} is very young compared to the history of merchant exchange, all applications and all consequences of this theory are not yet well evaluated, but all the technologies are available for a generalized primary usage. Our proposal opens a new and exciting field of research and investigation.
\begin{figure} \centering \scalebox{.75}{
\begin{tikzpicture}
\path[mindmap,concept color=gray,text=white]
    node[concept] {Marginal Cost\vspace{3mm} \\{\Large $M_c$}}
    [clockwise from=0]
    child[concept color=red] { node[concept] {{\bf Post-industrial}\\{\Large $M_c=0$}\\{\sc Intangible Good}}}
    child[concept color=blue] { node[concept] {{\bf Industrial}\\{\Large $M_c=\epsilon$}\\{\sc Tangible Good}}}
    child[concept color=orange] { node[concept] {{\bf Pre-industrial}\\{\Large $M_c\max{}$}\\{\sc Prototype}}};
\draw (-2.5,-2) node[text width=2cm, left]{\Large \color{gray}{Physical\\ World}};
\draw (4.6,2) node[left]{\Large \color{gray}{Internet}};
\draw (6,-5) node[left]{\color{blue}{\sc Copyright}};
\end{tikzpicture}
}
\caption{The Marginal Cost defines the type of economy; $M_c=0$ for {\sc ig}, $M_c=\epsilon$ for {\sc tg} and $M_c=\max{}$ for prototype.}
\label{fig_mindmap}
\end{figure}
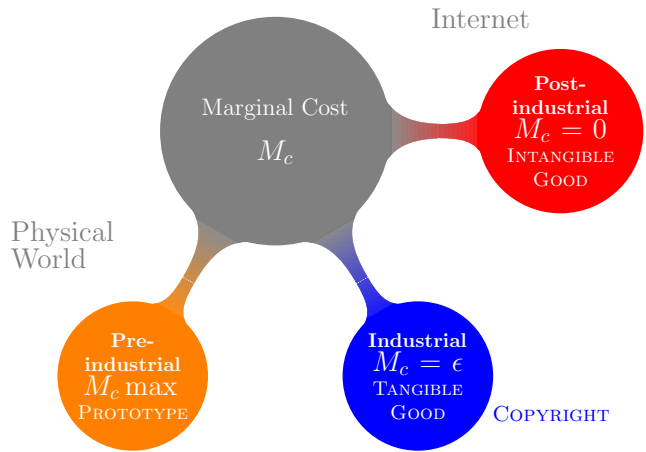

\section{Theory construction}

\newtheorem{hyp}{Hypothesis}
\newtheorem{axiom}{Axiom}
\newtheorem{defi}{Definition}
\newtheorem{theo}{Theorem}
\newtheorem{principle}{Principle}

Starting from the evaluation of {\em marginal cost} (Figure \ref{fig_mindmap}), we are proposing an axiomatization of economics for the {\em Internet}.

\begin{axiom}
An {\bf Intangible Good}\footnote{see \url{fr.wikipedia.org/wiki/Bien_immateriel} } ({\sc ig}) is a virtual object having a significant {\bf value} for a set of individuals, and a {\bf null} margin cost. 
\end{axiom}

Only the {\bf Internet} is able to save and to publish an {\sc ig}. Any file can be duplicated on any network node at no cost. End users are investing themselves in terminal computers, phones, storage devices, so the marginal cost for the producer of an {\sc ig} is null.   
Any tangible good ({\sc tg}) may have a very low margin cost with large scale mass production, but this cost is never null. 
Internet also store data for private communications, without any value in public publishing. This data is not considered as an {\sc ig}.

\begin{description}
\item[Creator:]
The {\em creator} of an {\sc ig} is one individual or a group of individuals using high skills and spending time to create the {\sc ig}. This work deserve a direct or indirect financial reward for the creator.
\item[Customer:]
A {\em customer} of an {\sc ig} in one individual owning a sufficient amount of money to acquire the right to {\em use} that {\sc ig} any time all his life, on any device\footnote{This right is referenced as the "mobiquity" right.}, without any {\sc drm}\footnote{Digital Right Management} attached nor advertising.
\item[Transaction:] (Figure \ref{fig_ig})
For a given {\sc ig}, Internet allows to define a {\sc one-to-many} temporal relation $\mathcal{F}_c(i,t)$ between the {\em creator} and the effective $i$ {\em customers}. As soon as the {\sc ig} is published, customers are free to choose the time for buying the {\sc ig}, without the creator agreement. 
In the same {\sc ig} relation, any new buyer, in position $i$, time $t$, may spend a {\em price} $\mathcal{P}_i^t$, making for the creator an {\em income} $\mathcal{I}_i^t$ and for the $i-1$ previous $j$ buyers as a {\em refund} $\mathcal{R}_{ij}^t$.
Unlike for a {\sc tg} transaction requiring transportation and transformations,
no intermediate actor is requested in the pure digital {\sc ig} transaction, thus no additional fee is required in {\sc ig} relation. So the following equation states:
$$
\forall i \in \mathbb{N}^* \hspace{1cm}
\mathcal{P}_i^t = \mathcal{I}_i^t + \sum_{j=1}^{j\leq i}{ \mathcal{R}^t_{ij}} 
$$
\end{description}
Starting from now, the time parameter $t$ is skipped because our main proposal is time independent, but we do not exclude to define in next developments a time dependent solution, specially for time valuated {\sc ig} like for flash paper news. 

\begin{hyp}
The price $\mathcal{P}_i$ is only dependent of the position $i$ of the buyer in the list (eventually the time of the buying action), but never dependent of buyer personal features/data or buyer financial capabilities.
\end{hyp}

\begin{hyp}
For a given {\sc ig} and knowing that the margin cost is null while the production cost is finite. It is a fair principle to bound the cumulative income $\mathcal{T}_i$ with a fixed value, $\mathcal{T}_{\infty}$, chosen by the creator and known universally. 
$$
\forall i \in \mathbb{N}^* \hspace{1cm} \mathcal{T}_i = \sum_{k=1}^{k\leq i}{\mathcal{I}_k} \hspace{1cm} \lim_{i \to \infty} \mathcal{T}_i = \mathcal{T}_{\infty}
$$
\end{hyp}
Then we have:
\begin{equation}
\lim_{i \to \infty} \mathcal{P}_i  = 0
\end{equation}
As $\mathcal{I}$ is always positive or null, the cummulative income $\mathcal{T}$ is increasing:
$$
\forall i \in \mathbb{N}^* \hspace{1cm}  \mathcal{T}^{'}_i \geq 0 
$$

\begin{hyp}
For a given {\sc ig}, the price function $\mathcal{P}_i$ is decreasing. If two customers ask to buy the same {\sc ig} at the same time, the displayed price has to be higher than the effective price.
$$
\forall i \in \mathbb{N}^* \hspace{1cm}  \mathcal{P}^{'}_i \leq 0 
$$
\end{hyp}

\begin{hyp}
For a given {\sc ig} and for any purchase number $i$, the refunding values $\mathcal{R}_{ij}$ for $j<i$ are equal.
$$
\forall i \in \mathbb{N}^* \hspace{1cm} \forall j < i \hspace{1cm} \mathcal{R}_{ij} = \mathcal{R}_{i}   
$$
\end{hyp}
Then we can verify the equations:

\begin{align}
\mathcal{P}_i &= \mathcal{I}_i + (i-1) \mathcal{R}_i \\ 
\mathcal{T}_i &= i \mathcal{P}_i \\
\mathcal{I}_i &= \mathcal{T}_{i} - \mathcal{T}_{i-1}
\end{align}

\begin{theo}
At the same time, all $i$ buyers had payed, including the refunds, the very same price to get the same {\sc ig}. This price is equal to $\mathcal{P}_i$. 
\end{theo}

\begin{theo}
For an {\sc ig} given by its creator with initial price $\mathcal{P}_1$ and a limit cumulative income $\mathcal{T}_{\infty}$, It exists a solution satisfying previous hypothesis. 
\end{theo}

The previous hypothesis can be summerized in the less formalized principle of {\sc fairness}:
\begin{principle}
Creator cumulative income of an {\sc ig} is bounded while every time, all buyers pay the very same price down to zero.
\end{principle}

The most significant relation is between unitary price and cumulative income and is called the {\em ``Pelinquin"} equation, showing the fragile equilibrium between increasing $\mathcal{T}$ and decreasing $\mathcal{P}$:
\begin{equation}
{\color{red}{\mathcal{T}_i^\nearrow{}=i\mathcal{P}_i^\searrow{}}} \tag{{\em ``Pelinquin"} {\tiny continuous} }
\end{equation}

We seen that {\sc ig} transactions follows rules not as simple as for the physical world where only a {\sc one-to-one} relation occurs. We used to use the  equation $\mathcal{P} = \mathcal{I} + \sum_i{f_i}$. The $f_i$ are fees taken by intermediaries for transportation and transformation and are subject to speculation. Prices, incomes are scalars and refund is null for tangibles goods. However, for {\sc ig}, its price is a function and the relation is a little more complex to manage the automatic refund. We introduced a breakthrough in traditional economics exchange.\\
One can note that the {\em ``pelinquin"} equation still applies for {\em pre-industrial} phase in the simple case where $i=1$, that is {$\color{blue}{\mathcal{T}=\mathcal{P}}$} and also for {\em industrial} phase where unitary price is constant; then {$\color{blue}{\mathcal{T}_i=i\mathcal{P}}$}.

The next section proposes a function solution family with "smooth" variations. This family requires only one tuning parameter $\xi \in [0,1]$ called speed parameter. 
\begin{figure} \centering
\tikzset{
  ig/.style={cloud, draw=none, text=white, fill=red!40, cloud puffs=10,cloud puff arc=120, aspect=2, inner ysep=-4mm},
  b/.style={rectangle,rounded corners=4pt,draw=none,text=black,fill=blue!15},
  tg/.style={rectangle,draw=none,text=white,fill=red!40},
  mya/.style={->, very thick,line width=3pt,draw=gray}, 
  myv/.style={->, very thick,line width=3pt,draw=black!40!green},
  myr/.style={->, thick,draw=gray, dashed},
}
\begin{tikzpicture}[align=center]
  \draw[->,very thick,draw=blue!50] (-1,2) -- (7,2) node[right] {time};
  \node[b] (v) at (0,0) {{\huge\Smiley\Mobilefone}\\Author(s)};
    \node[ig] (ig) at (2,2) {\bf \small {\sc Intangible}\\{\sc Good}};
    \node[b] (a1) at (1,4) {$1^{st}$ buyer\\{\huge\Smiley\Mobilefone}}; 
    \node[b] (a2) at (4,4) {$2^{nd}$ buyer\\{\huge\Smiley\Mobilefone}};
    \node[b] (a3) at (7.2,4) {\ldots buyer\\{\huge\Smiley\Mobilefone}};
    \draw[myv] (v) -- (ig);
    \draw[mya] (a1) -- (ig);\draw[mya] (a2) -- (ig);\draw[mya] (a3) -- (ig);
    \draw[myr] (a2) -- (a1);\draw[myr] (a3) -- (a2);
    \draw[myr] (a1) -- (v);\draw[myr] (a2) -- (v);\draw[myr] (a3) -- (v);
  \draw[-,very thick,draw=green!50] (0,1.7) -- (0,2.3) node[left] {\small Publishing};
\end{tikzpicture}
\caption{Intangible Good; a perpetual {\em many-to-many}, relation.}\label{fig_ig}
\end{figure}
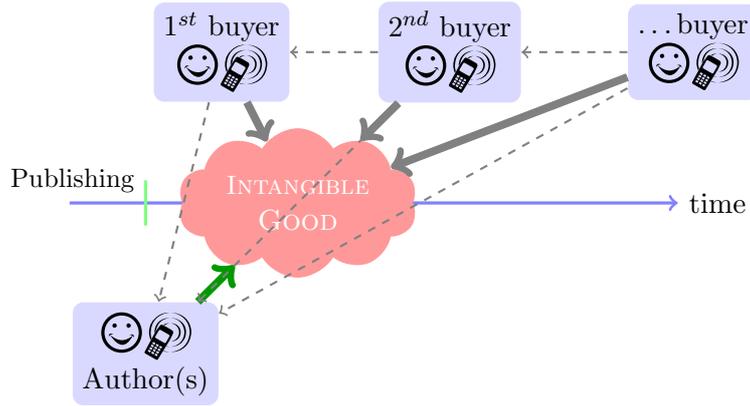

\begin{figure} \centering
\begin{tikzpicture} 
  \begin{axis}[ xlabel=$n$ (number of buyers), ylabel=$\sqcup$ (cup), width=11cm, height=7cm]
    \addplot[smooth,blue] plot coordinates { (1, 5.00) (2, 2.25) (3, 2.25) (4, 2.24) (5, 2.24) (6, 2.23) (7, 2.23) (8, 2.22) (9, 2.22) (10, 2.21) (11, 2.21) (12, 2.20) (13, 2.20) (14, 2.19) (15, 2.19) (16, 2.18) (17, 2.18) (18, 2.17) (19, 2.17) (20, 2.16) (21, 2.16) (22, 2.15) (23, 2.15) (24, 2.14) (25, 2.14) (26, 2.13) (27, 2.13) (28, 2.12) (29, 2.12) (30, 2.11)  };
    \addlegendentry{$\mathcal{I}$ncome}
    \addplot[smooth,color=red] plot coordinates { (1, 5.00) (2, 3.63) (3, 3.17) (4, 2.94) (5, 2.80) (6, 2.70) (7, 2.63) (8, 2.58) (9, 2.54) (10, 2.51) (11, 2.48) (12, 2.46) (13, 2.44) (14, 2.42) (15, 2.41) (16, 2.39) (17, 2.38) (18, 2.37) (19, 2.36) (20, 2.35) (21, 2.34) (22, 2.33) (23, 2.32) (24, 2.31) (25, 2.31) (26, 2.30) (27, 2.29) (28, 2.29) (29, 2.28) (30, 2.28)  };
    \addlegendentry{$\mathcal{P}$rice}
    \addplot[smooth,color=green] plot coordinates { (1, 0.00) (2, 1.37) (3, 0.46) (4, 0.23) (5, 0.14) (6, 0.09) (7, 0.07) (8, 0.05) (9, 0.04) (10, 0.03) (11, 0.03) (12, 0.02) (13, 0.02) (14, 0.02) (15, 0.02) (16, 0.01) (17, 0.01) (18, 0.01) (19, 0.01) (20, 0.01) (21, 0.01) (22, 0.01) (23, 0.01) (24, 0.01) (25, 0.01) (26, 0.01) (27, 0.01) (28, 0.01) (29, 0.01) (30, 0.01)  };
    \addlegendentry{$\mathcal{R}$efund}
  \end{axis}
\end{tikzpicture}
\caption{Income, Price and Refund values for a 5$\sqcup$1000 {\sc ig} and $\xi=0.3$.}
\label{fig_param}
\end{figure}
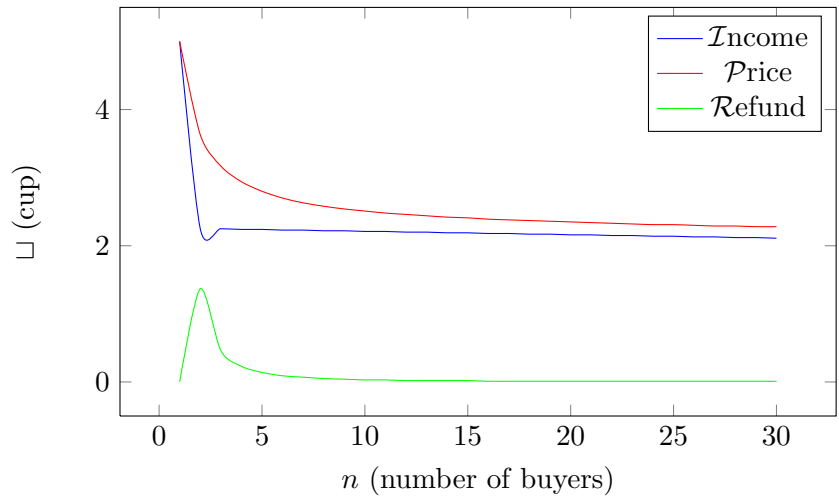

\begin{figure} \centering
\begin{tikzpicture}
  \begin{axis}[ xlabel=$n$ (number of buyers), ylabel=$\sqcup$ (cup), width=11cm, height=7cm]
    \addplot[smooth,blue] plot coordinates { (1,10.00) (2,12.61) (3,15.15) (4,17.61) (5,20.00) (6,22.32) (7,24.58) (8,26.76) (9,28.89) (10,30.95) (11,32.96) (12,34.90) (13,36.79) (14,38.62) (15,40.41) (16,42.13) (17,43.81) (18,45.44) (19,47.03) (20,48.56) (21,50.06) (22,51.51) (23,52.91) (24,54.28) (25,55.61) (26,56.89) (27,58.14) (28,59.36) (29,60.54) (30,61.68) (31,62.80) (32,63.87) (33,64.92) (34,65.94) (35,66.93) (36,67.89) (37,68.82) (38,69.73) (39,70.60) (40,71.46) (41,72.28) (42,73.09) (43,73.87) (44,74.63) (45,75.36) (46,76.08) (47,76.77) (48,77.45) (49,78.10) (50,78.74) (51,79.35) (52,79.95) (53,80.53) (54,81.10) (55,81.65) (56,82.18) (57,82.70) (58,83.20) (59,83.69) (60,84.16) (61,84.62) (62,85.07) (63,85.50) (64,85.92) (65,86.33) (66,86.73) (67,87.11) (68,87.48) (69,87.85) (70,88.20) (71,88.54) (72,88.88) (73,89.20) (74,89.51) (75,89.82) (76,90.11) (77,90.40) (78,90.68) (79,90.95) (80,91.21) (81,91.47) (82,91.71) (83,91.95) (84,92.19) (85,92.41) (86,92.63) (87,92.85) (88,93.05) (89,93.26) (90,93.45) (91,93.64) (92,93.83) (93,94.01) (94,94.18) (95,94.35) (96,94.51) (97,94.67) (98,94.83) (99,94.98) (100,95.12) (101,95.26) (102,95.40) (103,95.53) (104,95.66) (105,95.79) (106,95.91) (107,96.03) (108,96.15) (109,96.26) (110,96.37) (111,96.47) (112,96.57) (113,96.67) (114,96.77) (115,96.86) (116,96.95) (117,97.04) (118,97.13) (119,97.21) (120,97.29) (121,97.37) (122,97.45) (123,97.52) (124,97.59) (125,97.66) (126,97.73) (127,97.80) (128,97.86) (129,97.92) (130,97.98) (131,98.04) (132,98.10) (133,98.15) (134,98.21) (135,98.26) (136,98.31) (137,98.36) (138,98.41) (139,98.45) (140,98.50) (141,98.54) (142,98.58) (143,98.62) (144,98.66) (145,98.70) (146,98.74) (147,98.78) (148,98.81) (149,98.85) (150,98.88) (151,98.91) (152,98.95) (153,98.98) (154,99.01) (155,99.03) (156,99.06) (157,99.09) (158,99.12) (159,99.14) (160,99.17) (161,99.19) (162,99.21) (163,99.24) (164,99.26) (165,99.28) (166,99.30) (167,99.32) (168,99.34) (169,99.36) (170,99.38) (171,99.40) (172,99.41) (173,99.43) (174,99.45) (175,99.46) (176,99.48) (177,99.49) (178,99.51) (179,99.52) (180,99.54) (181,99.55) (182,99.56) (183,99.58) (184,99.59) (185,99.60) (186,99.61) (187,99.62) (188,99.63) (189,99.65) (190,99.66) (191,99.67) (192,99.68) (193,99.68) (194,99.69) (195,99.70) (196,99.71) (197,99.72) (198,99.73) (199,99.74) (200,99.74) (201,99.75) (202,99.76) (203,99.77) (204,99.77) (205,99.78) (206,99.78) (207,99.79) (208,99.80) (209,99.80) (210,99.81) (211,99.81) (212,99.82) (213,99.82) (214,99.83) (215,99.83) (216,99.84) (217,99.84) (218,99.85) (219,99.85) (220,99.86) (221,99.86) (222,99.87) (223,99.87) (224,99.87) (225,99.88) (226,99.88) (227,99.88) (228,99.89) (229,99.89) (230,99.89) (231,99.90) (232,99.90) (233,99.90) (234,99.91) (235,99.91) (236,99.91) (237,99.91) (238,99.92) (239,99.92) (240,99.92) (241,99.92) (242,99.93) (243,99.93) (244,99.93) (245,99.93) (246,99.93) (247,99.94) (248,99.94) (249,99.94) (250,99.94) (251,99.94) (252,99.94) (253,99.95) (254,99.95) (255,99.95) (256,99.95) (257,99.95) (258,99.95) (259,99.95) (260,99.96) (261,99.96) (262,99.96) (263,99.96) (264,99.96) (265,99.96) (266,99.96) (267,99.96) (268,99.97) (269,99.97) (270,99.97) (271,99.97) (272,99.97) (273,99.97) (274,99.97) (275,99.97) (276,99.97) (277,99.97) (278,99.97) (279,99.97) (280,99.98) (281,99.98) (282,99.98) (283,99.98) (284,99.98) (285,99.98) (286,99.98) (287,99.98) (288,99.98) (289,99.98) (290,99.98) (291,99.98) (292,99.98) (293,99.98) (294,99.98) (295,99.98) (296,99.98) (297,99.99) (298,99.99) (299,99.99)  };
    \addlegendentry{$\mathcal{T}$otal income}
    \addplot[smooth,color=red] plot coordinates { (1,10.00) (2, 6.31) (3, 5.05) (4, 4.40) (5, 4.00) (6, 3.72) (7, 3.51) (8, 3.35) (9, 3.21) (10, 3.10) (11, 3.00) (12, 2.91) (13, 2.83) (14, 2.76) (15, 2.69) (16, 2.63) (17, 2.58) (18, 2.52) (19, 2.48) (20, 2.43) (21, 2.38) (22, 2.34) (23, 2.30) (24, 2.26) (25, 2.22) (26, 2.19) (27, 2.15) (28, 2.12) (29, 2.09) (30, 2.06) (31, 2.03) (32, 2.00) (33, 1.97) (34, 1.94) (35, 1.91) (36, 1.89) (37, 1.86) (38, 1.83) (39, 1.81) (40, 1.79) (41, 1.76) (42, 1.74) (43, 1.72) (44, 1.70) (45, 1.67) (46, 1.65) (47, 1.63) (48, 1.61) (49, 1.59) (50, 1.57) (51, 1.56) (52, 1.54) (53, 1.52) (54, 1.50) (55, 1.48) (56, 1.47) (57, 1.45) (58, 1.43) (59, 1.42) (60, 1.40) (61, 1.39) (62, 1.37) (63, 1.36) (64, 1.34) (65, 1.33) (66, 1.31) (67, 1.30) (68, 1.29) (69, 1.27) (70, 1.26) (71, 1.25) (72, 1.23) (73, 1.22) (74, 1.21) (75, 1.20) (76, 1.19) (77, 1.17) (78, 1.16) (79, 1.15) (80, 1.14) (81, 1.13) (82, 1.12) (83, 1.11) (84, 1.10) (85, 1.09) (86, 1.08) (87, 1.07) (88, 1.06) (89, 1.05) (90, 1.04) (91, 1.03) (92, 1.02) (93, 1.01) (94, 1.00) (95, 0.99) (96, 0.98) (97, 0.98) (98, 0.97) (99, 0.96) (100, 0.95) (101, 0.94) (102, 0.94) (103, 0.93) (104, 0.92) (105, 0.91) (106, 0.90) (107, 0.90) (108, 0.89) (109, 0.88) (110, 0.88) (111, 0.87) (112, 0.86) (113, 0.86) (114, 0.85) (115, 0.84) (116, 0.84) (117, 0.83) (118, 0.82) (119, 0.82) (120, 0.81) (121, 0.80) (122, 0.80) (123, 0.79) (124, 0.79) (125, 0.78) (126, 0.78) (127, 0.77) (128, 0.76) (129, 0.76) (130, 0.75) (131, 0.75) (132, 0.74) (133, 0.74) (134, 0.73) (135, 0.73) (136, 0.72) (137, 0.72) (138, 0.71) (139, 0.71) (140, 0.70) (141, 0.70) (142, 0.69) (143, 0.69) (144, 0.69) (145, 0.68) (146, 0.68) (147, 0.67) (148, 0.67) (149, 0.66) (150, 0.66) (151, 0.66) (152, 0.65) (153, 0.65) (154, 0.64) (155, 0.64) (156, 0.64) (157, 0.63) (158, 0.63) (159, 0.62) (160, 0.62) (161, 0.62) (162, 0.61) (163, 0.61) (164, 0.61) (165, 0.60) (166, 0.60) (167, 0.59) (168, 0.59) (169, 0.59) (170, 0.58) (171, 0.58) (172, 0.58) (173, 0.57) (174, 0.57) (175, 0.57) (176, 0.57) (177, 0.56) (178, 0.56) (179, 0.56) (180, 0.55) (181, 0.55) (182, 0.55) (183, 0.54) (184, 0.54) (185, 0.54) (186, 0.54) (187, 0.53) (188, 0.53) (189, 0.53) (190, 0.52) (191, 0.52) (192, 0.52) (193, 0.52) (194, 0.51) (195, 0.51) (196, 0.51) (197, 0.51) (198, 0.50) (199, 0.50) (200, 0.50) (201, 0.50) (202, 0.49) (203, 0.49) (204, 0.49) (205, 0.49) (206, 0.48) (207, 0.48) (208, 0.48) (209, 0.48) (210, 0.48) (211, 0.47) (212, 0.47) (213, 0.47) (214, 0.47) (215, 0.46) (216, 0.46) (217, 0.46) (218, 0.46) (219, 0.46) (220, 0.45) (221, 0.45) (222, 0.45) (223, 0.45) (224, 0.45) (225, 0.44) (226, 0.44) (227, 0.44) (228, 0.44) (229, 0.44) (230, 0.43) (231, 0.43) (232, 0.43) (233, 0.43) (234, 0.43) (235, 0.43) (236, 0.42) (237, 0.42) (238, 0.42) (239, 0.42) (240, 0.42) (241, 0.41) (242, 0.41) (243, 0.41) (244, 0.41) (245, 0.41) (246, 0.41) (247, 0.40) (248, 0.40) (249, 0.40) (250, 0.40) (251, 0.40) (252, 0.40) (253, 0.40) (254, 0.39) (255, 0.39) (256, 0.39) (257, 0.39) (258, 0.39) (259, 0.39) (260, 0.38) (261, 0.38) (262, 0.38) (263, 0.38) (264, 0.38) (265, 0.38) (266, 0.38) (267, 0.37) (268, 0.37) (269, 0.37) (270, 0.37) (271, 0.37) (272, 0.37) (273, 0.37) (274, 0.36) (275, 0.36) (276, 0.36) (277, 0.36) (278, 0.36) (279, 0.36) (280, 0.36) (281, 0.36) (282, 0.35) (283, 0.35) (284, 0.35) (285, 0.35) (286, 0.35) (287, 0.35) (288, 0.35) (289, 0.35) (290, 0.34) (291, 0.34) (292, 0.34) (293, 0.34) (294, 0.34) (295, 0.34) (296, 0.34) (297, 0.34) (298, 0.34) (299, 0.33)  };
    \addlegendentry{$\mathcal{P}$rice}
  \end{axis}
\end{tikzpicture}
\caption{Income and Price evolution of a 1$\sqcup$100 {\sc ig} for $\xi=0.25$.}
\label{fig_exp}
\end{figure}
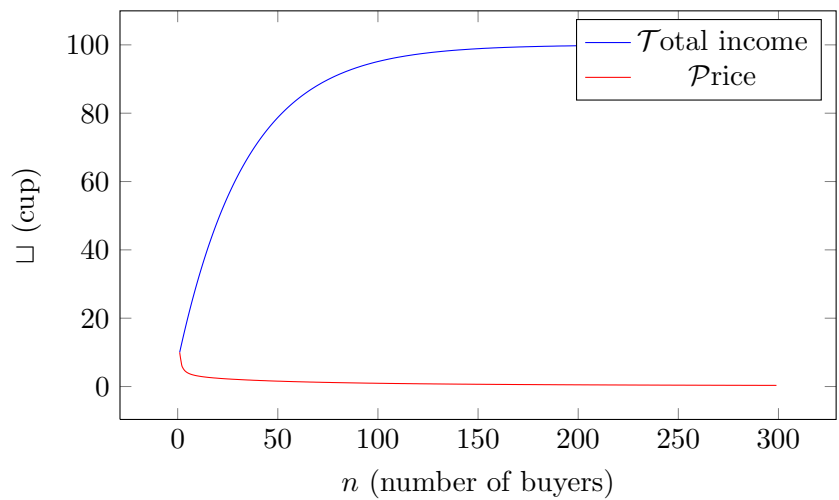

\section{The exponential family}
For an {\sc ig} sold at the first price $\mathcal{P}_1$ and expecting an income $\mathcal{T}_{\infty}$, the three related computed values are:
\begin{itemize}
\item $\mathcal{P}_i$ the price a buyer has to pay the good in position $i$
\item $\mathcal{I}_i$ the additional income to the creator for a purchase in position $i$, $\mathcal{T}_i$ is the cumulative income.
\item $\mathcal{R}_i$ the refunding value given to all previous ($i-1$) consumers of the same good
\end{itemize}
A piece linear model solution exists but we present here the exponential based solution. 
For this model, with a speed parameter $\xi \in [0,1]$ selected by the creator, and with:
$$
\lambda = \left( \frac{ \mathcal{T}_{\infty} - \mathcal{P}_1}{ \mathcal{T}_{\infty} - 2 \mathcal{P}_1} \right)^{\xi}
$$
we have $\mathcal{I}_1 = \mathcal{P}_1$, $\mathcal{R}_1 = 0$ and $\forall i > 1$ and ({\em Figure \ref{fig_param}}):
\begin{align}
\mathcal{P}_i &= \frac{ \mathcal{T}_{\infty} + (\mathcal{P}_1 - \mathcal{T}_{\infty})\lambda^{1-i}}{i} \\ 
\mathcal{I}_i &= (1-\lambda)(\mathcal{P}_1 - \mathcal{T}_{\infty})\lambda^{1-i} \hspace{4mm} \\
\mathcal{R}_i &= \frac{ \mathcal{T}_{\infty}+ \lambda^{1-i} \left( 1+i(\lambda-1) \right) (\mathcal{P}_1 - \mathcal{T}_{\infty})}{i(i-1)}
\end{align} 
The total income ({\em Figure \ref{fig_exp}}) is:
\begin{equation}
\mathcal{T}_i = \mathcal{T}_{\infty} + (\mathcal{P}_1 - \mathcal{T}_{\infty})\lambda^{1-i} 
\end{equation}
$\mathcal{P}_i$, $\mathcal{I}_i$ and $\mathcal{R}_i$ are decreasing while $\mathcal{T}_i$ is increasing and we check that:
$$
\lim_{i \to \infty} \mathcal{P}_i = 0 \hspace{1cm} \lim_{i \to \infty} \mathcal{T}_i = \mathcal{T}_{\infty}
$$
\subsection{The numerical rounding issue}
The implementation of the solution over the {\em Internet} raises a numerical rounding issue for the money. There is computational benefits to use integers instead of floating point. The {\em Pelinquin} equation becomes with integers:

\begin{equation}
\exists k \in [0,i] \hspace{1cm} \mathcal{T}_i=(i-k)\mathcal{P}_i + k\left(\mathcal{P}_i+1\right) \tag{{\em ``Pelinquin"} {\tiny discrete} }
\end{equation}
All buyers will not pay exactly the same price (including refunds) as for the continuous case, but the {\em non-equity} is bounded by 1 unit (arround 10 cents according to conversion rate described hereafter). We have first to select carefuly the $k$ distribution parameter so that any buyers is always asked to pay only once (she never have to add 1 unit later on). Thus, the real integer income is slightly overestimated compared to the continuous case. Second, we must check that the cumulative income remains increasing.\\ 
Our investigations defined the interger-based {\em function-price} algorithm (See algorithm \ref{algo1})\footnote{implementation is available on \url{https://github.com/pelinquin/pingpongcash/blob/master/node.py}} that returns $\mathcal{P}_i$ and the distribution parameter $k$.
\begin{algorithm}
\caption{{\sc function-price} $(\mathcal{P}_1, \mathcal{T}_{\infty}, \xi, i)$}
\begin{algorithmic}
\REQUIRE $\mathcal{P}_1>0 \wedge \mathcal{T}_{\infty} > 2\mathcal{P}_1 \wedge i > 0 \wedge \xi \in [0,1] $
\ENSURE $\mathcal{T}^{\nearrow{}} \wedge \mathcal{P}^{\searrow{}} \wedge k \in [0,i]$

\IF{$\xi = 0$}
 \STATE $\mathcal{P} = \textsc{integer}\left(\mathcal{P}_1/i\right)$
 \IF{$\mathcal{P} = 0$}
  \IF{$i < \mathcal{T}_{\infty}$}
   \RETURN $0, 0$
  \ELSE
   \RETURN $0, i-\mathcal{T}_{\infty}$
  \ENDIF
 \ELSE
  \RETURN $\mathcal{P}, i\left(\mathcal{P}+1\right)-\mathcal{P}_1$
 \ENDIF
\ENDIF
\STATE $\lambda\leftarrow \left(\frac{\mathcal{T}_{\infty}-\mathcal{P}_1}{\mathcal{T}_{\infty}-2\mathcal{P}_1}\right)^\xi$
\STATE $\mathcal{T}_A\leftarrow \mathcal{T}_{\infty}+\left(\mathcal{P}_1-\mathcal{T}_{\infty}\right)\lambda^{1-i}$
\STATE $\mathcal{P}\leftarrow \textsc{integer}\left(\frac{\mathcal{T}_A}{i}\right)$
\STATE $j\leftarrow i$
\STATE $k\leftarrow i(\mathcal{P}+1)-\textsc{round}(\mathcal{T}_A)$
\WHILE{\sc true}
\STATE $j\leftarrow j+1$
\STATE $\mathcal{T}_B\leftarrow \mathcal{T}_{\infty}+\left(\mathcal{P}_1-\mathcal{T}_{\infty}\right)\lambda^{1-j}$
\STATE $\mathcal{Q}\leftarrow \textsc{integer}\left(\frac{\mathcal{T}_B}{j}\right)$
\STATE $y\leftarrow j(\mathcal{Q}+1)-\textsc{round}(\mathcal{T}_B)$
\IF{$\mathcal{P} \neq \mathcal{Q}$}
\RETURN $\mathcal{P}, k$
\ENDIF
\IF{$k \geq y+i-j$}
\STATE $k\leftarrow y+i-j$
\ELSE
\RETURN $\mathcal{P}, k$
\ENDIF
\IF{$k < 0$}
\RETURN $\mathcal{P}, 0$
\ENDIF
\IF{$j-y = \mathcal{T}_{\infty}$}
\RETURN $\mathcal{P}, k$
\ENDIF
\ENDWHILE
\end{algorithmic}
\label{algo1}
\end{algorithm}

Simulation confirms than cumulative income is increasing and price always decreasing for any buyer (refunds are always positive). However, we are looking for a {\bf mathematical proof} of those two properties with such algorithm. Moreover, many loops can occur in the {\em while} statement, and we will investigate a shorter and more optimized version of the algorithm.

From a general viewpoint, we argue that the paradigm shifting in using a refunding mechanism requires to introduce a new {\bf money/currency}, noted with the {\sc squarecup} symbol: $\sqcup$\footnote{pronounced \textipa{/k2p/} }, dedicated to {\sc ig}, with main features:
\begin{itemize}
\item $\sqcup$ is vector based; $(\mathcal{P}_1,\mathcal{T}_{\infty},\xi)$ or function based but not scalar based unlike \euro{}, \$ or \pounds .
\item $\sqcup$ is an universal/international currency dedicated to {\em intangible goods} of the Internet, world-wide by construction.
\item $\sqcup$ is from design integer based, with a unit value around 10 cents; the smaller price to get a significant {\sc ig}. The exchange rate value is explained in section 5.
\item $\sqcup$ is not sensitive to speculation actions\ldots see section 6 for details on this point.  
\end{itemize}
For integer computation, the full transaction has to remain well balanced. As soon as the cumulative income approach the expected value, all entities (price, income, refund, become very small, so we must select the more suited rounding policy.
An admissible solution is to round first the income $\mathcal{I}_i$ and second the refund value $\mathcal{R}_i$, then compute the price as:
$$
\mathcal{P}_i = \mathcal{I}_i + (i-1) \mathcal{R}_i 
$$
This way, all the money given by a new coming customer is shared between the creator and the previous buyers.\\

We may find two interesting values:
\begin{description}
\item[No refunding threshold]: the value $i_{nr}$ for which: $\forall i \ge i_{nr} \hspace{1cm} \mathcal{R}_i = 0 $ 
\item[Public domain threshold]: the value $i_{pd}$ for which: $\forall i \ge i_{pd} \hspace{1cm} \mathcal{I}_i = \mathcal{P}_i = \mathcal{R}_i = 0 $ 
\end{description}
It is obviously verified that $i_{nr} < i_{pd}$ and
as the $\sqcup$ money is integer based, we have:
\begin{equation} 
i_{pd} = \mathcal{T}_{\infty}
\end{equation}
The first $i_{pd}$ customers had paid $1\sqcup$ and all the other get the {\sc ig} for free.

\section{No paywall}
Beyond the idea that {\em Internet} pushing and speeding-up trading of {\em tangibles goods}, it remains today a {\em paywall} on the Net. The current main payment systems; {\em VISA, Mastercard, PayPal\ldots} are in the same time too complex, poorly secure, and very expensive for any citizen (merchant or customer). This is particularly unfortunate for {\sc ig} creators who can't publish their work directly on the Net, on their own server, just because they do not have access to an automatic, in the Net stack, free payment system to get their incomes.
{\em One click} payment attends to fix that paywall but raises privacy issues as we see for {\em Google-wallet} or {\em iTunesStore}. 
We argue that a distributed, open-source, free of charge solution is technically possible and it would promote a new peer-to-peer publishing, instead of concentration on huge intermediary private platforms. This digital payment system called {Ping-Pong-Cash} is fully adapted to $\sqcup$ trading for {\sc ig} but also provides great usage for traditional trading of {\sc tg}, using \euro{} or \$ currencies.\\
Authentication is a key point for digital payment. That's why our system requires a three ways authentication:
\begin{itemize}
\item Something you carry (phone, card, usb stick\ldots)
\item Something you know (PIN, passphrase\ldots)
\item Something on you (bio-metric data)
\end{itemize}
The smartphone is likely the best device to enable this strong autentication. It store a locally generated private key while the public key is readable universally on a Ditributed Hash Table {\sc dht}. A transaction is simply a message digitally signed by the buyer. The selected digital signature algorithm is {\sc ecdsa} with the {\em P521} NIST eliptic curve\footnote{\url{http://www.nsa.gov/ia/_files/nist-routines.pdf}}. Private key usage is protected by {\sc aes 256} symetric encryption.
Any user who buy an {\sc ig} received an encrypted {\sc url}, when decrypted, downloads the full file.

\section{Exchange rate proposal}
The $\sqcup$ money is not created from scratch, it is more considered as a {\em unit} computed value based on a shared formula. The $\sqcup$ currency is then convertible with most local currencies in the World.   
Let define as $\cal{C}$ the finite subset of $n$ currencies not including the $\sqcup$. $r_{ij}^t$ is the exchange rate between the currency $i$ and currency $j$ at discrete date $t$. Each currency $i$ has a known volume $v_i$ in the world and $v$ is the total volume. Volumes are supposed stable during long time periods.
The trivial solution for fixing $r_{\sqcup{}k}^t$ for any currency $k$ would have been to sum all current exchange rates for currencies in $\mathcal{C}$ weighted by their respective volume $v_i$ as:
$$
\forall t \in \mathbb{N}, \forall k \in \mathcal{C} \hspace{1cm}  r_{\sqcup{}k}^t = \frac{1}{v} \sum_{i=0}^{i<n} v_i r_{ik}^t
$$

As it as been set for the ECU in 1999\cite{ecu} for introducing the Euro. This definition would not minimize the time variation of the exchange rate with other currency, making an opportunity to engage speculation with such currency. To fix this issue, we consider that exchange rates are given each day, so $t'$ means the day after $t$.
Let select randomly a currency $k$ in $\cal{C}$ and we are facing the problem of defining the exchange rate $r_{\sqcup{}k}$ between $\sqcup$ and $k$.
One has to select an initial value $r_0 = r_{\sqcup{}k}^0$ at the date of birth for the international currency.

We proposes a recursive algorithm that computes $r_{\sqcup{}k}^{t'}$ knowing $r_{\sqcup{}k}^t$ for a given currency $k$ but all other exchange rates with other currencies $i$ are immediately computed with:
$$ 
r_{\sqcup{}i}^t = r_{\sqcup{}k}^t r_{ki}^t \hspace{1cm} \forall t \in \mathbb{N} \hspace{1cm} \forall i \in \mathcal{C} 
$$

Let define the value:

$$ 
\mathcal{V} = \min_{r_{\sqcup{}k}^{t'}}{\left( \sum_{i=0}^{i<n}{v_i\left|1 - \frac{r_{\sqcup{}i}^{t'}}{r_{\sqcup{}i}^t}\right|} \right)} 
$$

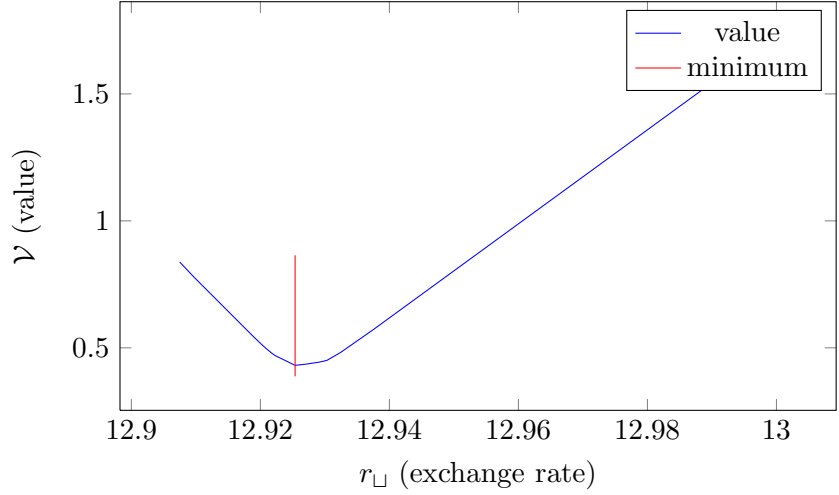
\begin{figure} \centering
\begin{tikzpicture}
  \begin{axis}[ xlabel=$r_{\sqcup}$ (exchange rate), ylabel=$\mathcal{V}$ (value), width=11cm, height=7cm]
    \addplot[blue] plot coordinates { (12.90749133230059, 0.8379940973829588) (12.90978046977524, 0.7766024295823113) (12.918891621109577, 0.5449568912452718) (12.920183409735028, 0.5135138340684771) (12.920844853556565, 0.49823293318453543) (12.92169738461423, 0.48038493948198746) (12.922202880131113, 0.4707411166522537) (12.922281645070745, 0.4695553989015693) (12.922283960509999, 0.4695237678373565) (12.92538187429505, 0.43103911785128457) (12.926951103757919, 0.4353406875089234) (12.929164004474131, 0.4438032785665863) (12.9298508207566, 0.4475984788870788) (12.930339090199137, 0.4509762849925367) (12.93241355789525, 0.4813706874563142) (12.937508979562645, 0.5717872476157182) (13, 1.7289717934780378) };
    \addlegendentry{value}
    \addplot[smooth,color=red] plot coordinates { (12.92538187429505,0.3879352060661561) (12.92538187429505,0.8644858967390189) };
    \addlegendentry{minimum}
  \end{axis}
\end{tikzpicture}
\caption{Value function for $r_{\sqcup{}\${}}^t$ = 13, then minimum reach for $r_{\sqcup{}\${}}^{t'}$ = 12.9253818743.}
\label{fig_plot}
\end{figure}

This value $\cal{V}$ can be written:
$$
\mathcal{V} = \min_{r_{\sqcup{}k}^{t'}}{\left( \sum_{i=0}^{i<n}{v_i\left|1 - \frac{r_{ki}^{t'}}{r_{ki}^tr_{\sqcup{}k}^t} r_{\sqcup{}k}^{t'} \right|} \right)} = \min_{r_{\sqcup{}k}^{t'}}{\left( \sum_{i=0}^{i<n}{\left|a_i r_{\sqcup{}k}^{t'} + b_i \right|} \right)}
$$
with $a_i$ and $b_i$ known constants.\\
The previous multi pieces linear equation has one solution ({Figure \ref{fig_plot}}) satisfying: $\exists j \in \mathcal{C}, r_{\sqcup{}k}^{t'} = -b_j/a_j$
For such currency $j$ that minimize the value $\cal{V}$, we have:
$$
r_{\sqcup{}k}^{t'} = \frac{r_{kj}^t}{r_{kj}^{t'}}r_{\sqcup{}k}^t
$$
If:
$$
\mathcal{V}_i = \sum_{i\ne{}j}{v_i\left|1 - \frac{r_{\sqcup{}i}^{t'}}{r_{\sqcup{}i}^t}\right|}
$$
then $j$ is such that $\mathcal{V} = \min{(\mathcal{V}_i)}$\\
The algorithm simply computes the $\mathcal{V}_i$ for all currencies of $\cal{C}$ and find the $j$ currency that does not change its rate between $t$ and $t'$.
The currency solution $j$ is always independent of the selected currency $k$ used at first reference. 
\\
Let now define the value function:
$$
\mathcal{V} = \sum_{i=0}^{i<n}{v_i\left(1 - \frac{r_{\sqcup{}i}^{t'}}{r_{\sqcup{}i}^t}\right)^2}
$$
This polynomial function of degree two has a unique minimum for:
$$
\forall k \in \mathcal{C} \hspace{1cm}
r_{\sqcup{}k}^{t'} =  
\left( 
\frac{ \displaystyle \sum_{i=0}^{i<n} { \frac{r_{ki}^{t'}}{r_{ki}^t}  v_i          }}
     { \displaystyle \sum_{i=0}^{i<n} {\left( \frac{r_{ki}^{t'}}{r_{ki}^t} \right) ^2 \! v_i   }} 
\right) r_{\sqcup{}k}^{t} 
= \mathcal{K}_k^{tt'} . r_{\sqcup{}k}^{t}
$$
Knowing exchange rates between all currencies in $\mathcal{C}$ for all days from an initial date, one can compute the $\mathcal{K}_k^{tt'}$ delta coefficient and then exchange rate with $\sqcup$ for all those days, starting with a fixed initial value. 
Our proposal for the most used currency in the set:
$$
\mathcal{C} = \{ \textsc{usd}, \textsc{eur}, \textsc{jpy}, \textsc{gsp}, \textsc{aud}, \textsc{chf}, \textsc{cad}, \textsc{hkd}, \textsc{sek}, \textsc{nzd},\textsc{sgd}, \textsc{krw}, \textsc{nok}, \textsc{mxn}, \textsc{inr} \}
$$ 
with their respective volume value\cite{wikiforex}:
$$
v = \{ 849, 391, 190, 129, 76, 64, 53, 24, 22, 16, 15, 15, 13, 13, 9 \}
$$

\section{Speculation}
\begin{figure} \centering
\tikzset{
  ig/.style={cloud, draw=none, text=white, fill=red!40, cloud puffs=10,cloud puff arc=120, aspect=2, inner ysep=-4mm},
  b/.style={rectangle,rounded corners=4pt,draw=none,text=black,fill=blue!15},
  tg/.style={rectangle,draw=none,text=white,fill=red!40},
  mya/.style={->, very thick,line width=2pt,draw=gray}, 
  myv/.style={->, very thick,line width=2pt,draw=black!40!green},
  myr/.style={->, very thick,line width=2pt,draw=black!40!black},
}
\begin{tikzpicture}
    \node[ig] (ig) at (2,1) {\bf {\sc ig/$\sqcup$}};
    \node[b] (a1) at (0,1) {{\huge\Gentsroom}}; 
    \node[b] (a2) at (4,1) {{\huge\Gentsroom}};
    \node[b] (tg) at (8,1.5) {\bf {\sc tg}};
    \node[b] (e) at (8,0.5) {\bf \euro };
    \node[b] (a3) at (6,1) {{\huge\Gentsroom}}; 
    \node[b] (a4) at (10,1) {{\huge\Gentsroom}};
    \draw[myv] (ig) -- (a1); \draw[myv] (ig) -- (a2);
    \draw[myr] (a3) -- (tg); \draw[myr] (tg) -- (a4);
    \draw[myv] (e) -- (a3); \draw[myv] (a4) -- (e);
\end{tikzpicture}
\caption{Trading difference between Intangible Good ({\sc ig}) and Tangible Good ({\sc tg}).}
\label{fig_trading}
\end{figure}
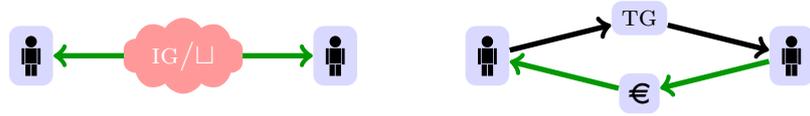

The exchange rate for $\sqcup$ as defined in the previous section protects the currency from speculation attacks. There are mainly five reasons for that resistance:
\begin{itemize}
\item As an international currency, available all over the Internet, there is no need for currency exchange. The ratio is the most stable from the set $\cal{C}$ of reference currencies, so any fluctuation of $\sqcup$ means that other currencies has changed. However, the algorithm is fully deterministic knowing the daily exchange rates between currencies in $\cal{C}$ and one exchange with $\sqcup$ the day before. Authorities or human evaluation cannot change the new exchange rate. Anybody in the world can compute it to get the same result\footnote{$\sqcup$ exchange rate started January 1, 2014 with an initial value $1\sqcup = 0.1$\euro. On May 7, 2014, the value is $1\sqcup = 0.09840473135599744$\euro. }. 
\item Any {\em Intangible Good} is sold on the Internet in $\sqcup$ at the same price whatever the development level of the population. Then, there is no local market places with different prices and possible speculation gains.
\item $\sqcup$ exchange with other currencies are subject to support a selling and buying tax going to the government or financial authority in charge of the foreign currency. For instance to exchange \$ into \euro{} using $\sqcup$, one has to pay a tax to {\sc eec} for buying $\sqcup$ from \euro{} and a tax for the USA for selling $\sqcup$ into \$. As we see, there is no advantage for traders to use $\sqcup$ between currencies on the market exchange.   
\item $\sqcup$ is dedicated to {\em Intangible Good} exchange and the transaction is atomic (money in $\sqcup$ against the URL of the good). There is not a double transaction like in the tangible world ({Figure \ref{fig_trading}}). Any undervalued or overvalued transaction using $\sqcup$ is immediately suspect and may hide another illegal transaction. Transactions without a real intangible good are only allowed for the same individual to change local currency into $\sqcup$ in order to buy cultural goods or for an artist earning $\sqcup$ to change them into the local currency. It is not allowed for one person to $\sqcup$ from another person against euros for instance. There should always exists an intangible good published over the Internet by the seller.   
\item $\sqcup$ economics follows different rules than the classical economics for tangible goods. The mechanism of refunding of previous buyers when new buyers are coming makes $\sqcup$ the first functional or non scalar currency. Those different rules makes more clear the necessity to define a new currency. It could have been possible to computes transaction following the fair intangible good principle using euro, but then people would have to carefully make the difference between euros for tangible goods, without refunding and euros for intangible goods with refunding and bound income. Some sellers of cultural goods on the Internet may have use the same currency name to sell without refunding and without bound income, an clearly unfair transaction.
\end{itemize}

\section{Conclusion}
In {\em economics}, this paper proposed a new paradigm dedicated to {\em Intangible Good} {\sc trading} fully consistent with {\sc sharing}. Thanks to the {\em Internet} to make the theory possible and verifiable in real life. We introduced the $\sqcup$ convertible, functional money, we shown the resistance against speculation and we argued on the need to deploy as soon as possible, an easy-to-use, free-of-charge, open-source, distributed, secure digital payment system for person-to-person exchanges, both for {\sc ig} and for {\sc tg}. Next study will detail the network architecture and the peer-to-peer node design, including cryptographic primitives.

\end{document}